Score Predictor Factor Analysis as model for the identification of single-item indicators


André Beauducel & Norbert Hilger

University of Bonn, Bonn, Germany



Address for correspondence:

André Beauducel

Institute of Psychology

Rheinische Friedrich-Wilhelms-Universität Bonn

Kaiser-Karl-Ring 9

53111 Bonn

beauducel@uni-bonn.de




Abstract

Score Predictor Factor Analysis (SPFA) was introduced as a method that allows to compute factor score predictors that are -under some conditions- more highly correlated with the common factors resulting from factor analysis than the factor score predictors computed from the common factor model. In the present study, we investigate SPFA as a model in its own rights. In order to provide a basis for this, the properties and the utility of SPFA factor score predictors and the possibility to identify single-item indicators in SPFA loading matrices were investigated. Regarding the factor score predictors, the main result is that the best linear predictor of the score predictor factor analysis has not only perfect determinacy but is also correlation preserving. Regarding the SPFA loadings it was found in a simulation study that five or more population factors that are represented by only one variable with a rather substantial loading can more accurately be identified by means of SPFA than with conventional factor analysis. Moreover, the percentage of correctly identified single-item indicators was substantially larger for SPFA than for the common factor model. It is therefore argued that SPFA is a tool that can be especially helpful when very short scales or single-item indicators are to be identified.

Keywords: Factor analysis, Score Predictor Factor Analysis, determinacy, short scales, single-item indicator



Beauducel and Hilger (2019) proposed Score Predictor Factor Analysis (SPFA) in order to overcome the difference between the covariances reproduced by the factor loadings of the common factor model (CFM) and the covariances reproduced by the model implied by the CFM factor score predictors. The difference between the covariances reproduced by the CFM loadings and the covariances reproduced by the CFM factor score predictor is described in Beauducel (2007) and conditions where this difference is a minimum are investigated in Beauducel and Hilger (2015). Nevertheless, it might be regarded as a problem that the interpretation of CFM loadings and the interpretation of the loadings resulting from CFM factor score predictors can diverge (Beauducel, 2005). Therefore, the aim of SPFA is to assure that the SPFA loadings yield the same interpretation as the loadings that are implied by the SPFA factor score predictors.

The focus of Beauducel and Hilger (2019) was on the differences between the common factor model, principal component analysis, and SPFA, so that they investigated how well common factors can be recovered by SPFA factor score predictors and principal components. Accordingly, they investigated the correlation of the SPFA best linear predictors with the corresponding factors of the common factor model. They showed that -under some circumstances- the SPFA best linear predictor has larger correlations with the corresponding CFM factor than the CFM best linear factor score predictor itself. It might therefore be considered to compute the best linear SPFA factor score predictor as a substitute for the conventional CFM best linear predictor. Although the SPFA best linear predictor may be regarded as a substitute for the CFM best linear predictor, SPFA might also be considered as a model in its own right. In order to provide a basis for this, the properties and the possible utility of SPFA as a model for multivariate data analysis is investigated in the present study.

Using SPFA in its own rights implies that the validity of SPFA factor score predictors as indicators of the SPFA factors should be investigated. This aspect has not been addressed in Beauducel and Hilger (2019) as they focused on SPFA factor score predictors as indicators for CFM factors. Therefore, the algebraic description of SPFA factor score predictors and their validity for the SPFA factors will be addressed here. Typically, three aspects of the validity of factor score predictors are discussed: The correlation of the factor score predictors



with the factors, which is the determinacy of the factor score predictor (Guttman, 1955; Grice, 2001; Nicewander, 2019) and sometimes also termed 'validity' (Gorsuch, 1983), the similarity of the intercorrelation of the factor score predictors with the intercorrelation of the factors (correlation preserving), and conditionally unbiasedness, which implies that the factor score predictors have substantial correlations only with the intended factors. Whereas the correlation of the factor score predictors with the factors can be regarded as a form of convergent validity, conditional unbiasedness could be regarded as a form of discriminant validity. Structural similarity, a fourth criterion for the validity of factor score predictors, implies that the covariances (i.e., the non-diagonal elements of the covariance matrix) reproduced from the factor score predictor are as similar as possible to the covariances reproduced from the factors (Beauducel & Hilger, 2015). Thus, the algebraic investigation of SPFA factor score predictors will comprise these four aspects of score validity (determinacy, preserving inter-correlations, conditional unbiasedness, structural similarity).

Investigating SPFA as a model in its own rights also implies that the SPFA loadings are considered because, usually, the loadings are inspected as a basis for an interpretation of factors. According to Beauducel and Hilger (2015) a loading matrix containing only zero loadings and a single unit-loading per factor leads to a perfect fit of the model reproduced from the factor score predictors. Since SPFA is a model based on the factor score predictors, this implies that SPFA should be especially suitable for the analysis of loading patterns that are based on a few large loadings per factor and a large number of very small loadings. This is in line with the results of Beauducel and Hilger (2019), who found the largest correlation of SPFA factors with CFM factors when there was a single, very large salient loading per factor. Models with very few, very large loadings per factor occur when a single measured variable is used as an optimal indicator of a latent variable. This is relevant when very short scales or single-item indicators are required for large scale surveys (Rammstedt & Beierlein, 2014) and in research contexts where a very large number of constructs has to be represented (Ziegler, Kemper, & Kruyen, 2014). Moreover, Bergkvist and Rossiter (2007) demonstrate that the predictive validity of single-item measures can be equal to the predictive validity of multiple-item measures. Accordingly, the use of single-item measures might also be justified because it



can be more effective. Even though in many settings multiple-item measures are probably more valid than single-item measures, finding single-item measures with maximal validity is probably still an issue. According to Beauducel and Hilger (2015, 2019) one might expect that SPFA is especially useful for the identification of factors with very few large loadings so that SPFA could be used for the identification of very short scales or single-item measures. Therefore, the second aim of the present study is to compare SPFA loadings and CFM loadings for CFM population models based on factors with a very small number of variables with rather high loadings. Since the identification of the population loading patterns by means of CFM and SPFA might depend on factor rotation, different methods of factor rotation will also be considered. This aspect will be addressed by means of a simulation study where samples are drawn from the population.

In sum, the aim of the present study is to investigate the usefulness of SPFA as a model in its own rights. Accordingly, some definitions will be given before the validity of SPFA factor score predictors as predictors of SPFA factors will be described algebraically. Finally, as this could be an interesting application of SPFA, the identification of population models with a single large loading per factor and the selection of the variable with the largest population loading by means of SPFA and CFM will be compared by means of a simulation study.

## Definitions

The common factor model states that a random vector $\mathbf{x}$ of $p$ observed variables is predicted by a random vector $\mathbf{f}$ of $q$ common factors, with $E(\mathbf{ff}') = \mathbf{\Phi}$ and $diag(\mathbf{\Phi}) = \mathbf{I}$ and by a random vector $\mathbf{u}$ of $p$ uncorrelated unique factors, with $E(\mathbf{uu}') = \mathbf{I}$. There is a $p \times q$ loading matrix $\mathbf{\Lambda}$ containing the weights of the common factors and a $p \times p$ nonsingular diagonal matrix $\mathbf{\Psi}$ containing the non-zero weights of the unique factors, so that

$$\mathbf{x} = \mathbf{\Lambda f} + \mathbf{\Psi u}, \tag{1}$$

and

$$E(\mathbf{xx}') = \mathbf{\Sigma} = \mathbf{\Lambda \Phi \Lambda}' + \mathbf{\Psi}^2. \tag{2}$$

An additional matrix representing the residual covariances due to model error of the CFM may be added to the right hand side of Equation 2 (MacCallum & Tucker, 1991). In



exploratory factor analysis, the loadings of orthogonal factors are estimated in a first step. According to the Minres-method the residuals of the observed correlations should be a minimum (Harman & Jones, 1966). Accordingly, the Minres-loadings $\hat{\mathbf{\Lambda}}_M$ are estimated in order to minimize the residuals of the non-diagonal elements of the observed sample covariance matrix $\mathbf{S}$, that is

$$tr[(\mathbf{S}-\hat{\mathbf{\Lambda}}_M\hat{\mathbf{\Lambda}}_M^{'}-diag(\mathbf{S}-\hat{\mathbf{\Lambda}}_M\hat{\mathbf{\Lambda}}_M^{'}))^{'}(\mathbf{S}-\hat{\mathbf{\Lambda}}_M\hat{\mathbf{\Lambda}}_M^{'}-diag(\mathbf{S}-\hat{\mathbf{\Lambda}}_M\hat{\mathbf{\Lambda}}_M^{'}))]=\min. \qquad (3)$$

Similarly, SPFA loadings are estimated by minimizing the residuals

$$\begin{aligned} tr[(\mathbf{S}-\hat{\mathbf{\Lambda}}_{os}(\hat{\mathbf{\Lambda}}_{os}^{'}\mathbf{S}^{-1}\hat{\mathbf{\Lambda}}_{os})^{-1}\hat{\mathbf{\Lambda}}_{os}^{'}-diag(\mathbf{S}-\hat{\mathbf{\Lambda}}_{os}(\hat{\mathbf{\Lambda}}_{os}^{'}\mathbf{S}^{-1}\hat{\mathbf{\Lambda}}_{os})^{-1}\hat{\mathbf{\Lambda}}_{os}^{'}))^{'} \\ (\mathbf{S}-\hat{\mathbf{\Lambda}}_{os}(\hat{\mathbf{\Lambda}}_{os}^{'}\mathbf{S}^{-1}\hat{\mathbf{\Lambda}}_{os})^{-1}\hat{\mathbf{\Lambda}}_{os}^{'}-diag(\mathbf{S}-\hat{\mathbf{\Lambda}}_{os}(\hat{\mathbf{\Lambda}}_{os}^{'}\mathbf{S}^{-1}\hat{\mathbf{\Lambda}}_{os})^{-1}\hat{\mathbf{\Lambda}}_{os}^{'}))]=\min, \end{aligned} \qquad (4)$$

resulting in the orthogonal SPFA loadings $\hat{\mathbf{\Lambda}}_s=\hat{\mathbf{\Lambda}}_{os}(\hat{\mathbf{\Lambda}}_{os}^{'}\mathbf{S}^{-1}\hat{\mathbf{\Lambda}}_{os})^{-1/2}$, where "$-1/2$" denotes the inverse of the symmetric square root. As noted in Harman (1976), the corresponding oblique loading pattern can be computed as

$$\hat{\mathbf{\Lambda}}_{sp}=\hat{\mathbf{\Lambda}}_s(\mathbf{T}^{'})^{-1}=\hat{\mathbf{\Lambda}}_{os}(\hat{\mathbf{\Lambda}}_{os}^{'}\mathbf{S}^{-1}\hat{\mathbf{\Lambda}}_{os})^{-1/2}(\mathbf{T}^{'})^{-1}, \quad \text{with } \mathbf{T}^{'}\mathbf{T}=E(\mathbf{f}_s\mathbf{f}_s^{'})=\hat{\mathbf{\Phi}}_s, \qquad (5)$$

where $\mathbf{T}$ is a transformation matrix that is obtained by means of factor rotation. Minimizing Equation 4 yields

$$\mathbf{S}=\hat{\mathbf{\Lambda}}_s(\mathbf{T}^{'})^{-1}\hat{\mathbf{\Phi}}_s\mathbf{T}^{-1}\hat{\mathbf{\Lambda}}_s+\mathbf{\Psi}_s^2+\mathbf{\Omega}_s, \qquad (6)$$

where $\mathbf{\Omega}_s$ represents the residual correlations due to model error of the SPFA.

## SPFA factor score predictors

Inserting $\hat{\mathbf{\Lambda}}_{os}(\hat{\mathbf{\Lambda}}_{os}^{'}\mathbf{S}^{-1}\hat{\mathbf{\Lambda}}_{os})^{-1/2}(\mathbf{T}^{'})^{-1}$ for $\mathbf{\Lambda}$, $\mathbf{\Phi}_s$ for $\mathbf{\Phi}$, and $\mathbf{S}$ for $\mathbf{\Sigma}$ into $\mathbf{f}_{BL}=\mathbf{\Phi}\mathbf{\Lambda}^{'}\mathbf{\Sigma}^{-1}\mathbf{x}$ for the CFM best linear predictor (i.e. the regression score predictor; Thurstone, 1935) yields

$$\hat{\mathbf{f}}_{sBL}=\hat{\mathbf{\Phi}}_s\mathbf{T}^{-1}(\hat{\mathbf{\Lambda}}_{os}^{'}\mathbf{S}^{-1}\hat{\mathbf{\Lambda}}_{os})^{-1/2}\hat{\mathbf{\Lambda}}_{os}^{'}\mathbf{S}^{-1}\mathbf{x}, \qquad (7)$$

the SPFA best linear predictor, with

$$Cov(\hat{\mathbf{f}}_{sBL},\hat{\mathbf{f}}_{sBL})=\hat{\mathbf{\Phi}}_s\mathbf{T}^{-1}(\hat{\mathbf{\Lambda}}_{os}^{'}\mathbf{S}^{-1}\hat{\mathbf{\Lambda}}_{os})^{-1/2}\hat{\mathbf{\Lambda}}_{os}^{'}\mathbf{S}^{-1}\hat{\mathbf{\Lambda}}_{os}(\hat{\mathbf{\Lambda}}_{os}^{'}\mathbf{S}^{-1}\hat{\mathbf{\Lambda}}_{os})^{-1/2}(\mathbf{T}^{'})^{-1}\hat{\mathbf{\Phi}}_s=\hat{\mathbf{\Phi}}_s, \qquad (8)$$

so that $Var(\hat{\mathbf{f}}_{sBL})=diag(\hat{\mathbf{\Phi}}_s)=\mathbf{I}$. Determinacy, i.e., the correlation of $\hat{\mathbf{f}}_{sBL}$ with the SPFA factors $\mathbf{f}_s$ is therefore

$$Cor(\hat{\mathbf{f}}_{sBL},\mathbf{f}_s)=\hat{\mathbf{\Phi}}_s\mathbf{T}^{-1}(\hat{\mathbf{\Lambda}}_{os}^{'}\mathbf{S}^{-1}\hat{\mathbf{\Lambda}}_{os})^{-1/2}\hat{\mathbf{\Lambda}}_{os}^{'}\mathbf{S}^{-1}\hat{\mathbf{\Lambda}}_{os}(\hat{\mathbf{\Lambda}}_{os}^{'}\mathbf{S}^{-1}\hat{\mathbf{\Lambda}}_{os})^{-1/2}(\mathbf{T}^{'})^{-1}\hat{\mathbf{\Phi}}_s=\hat{\mathbf{\Phi}}_s. \qquad (9)$$

Equation 9 implies that $\hat{\mathbf{f}}_{sBL}$ has a perfect determinacy and is correlation-preserving. With respect to conditional unbiasedness it should be noted that the SPFA best linear factor score predictor correlates to the same degree with other factors than the factors themselves do.



Structural similarity of the SPFA factor score predictor means that the non-diagonal elements of the covariance matrix reproduced by the SPFA factors ($\hat{\Sigma}_{SPFA}$) are the same as the non-diagonal elements of the SPFA best linear factor score predictor ($\hat{\Sigma}_{sBL}$). This condition is given by Theorem 1.

**Theorem 1.** $\hat{\Sigma}_{SPFA} - diag(\hat{\Sigma}_{SPFA}) = \hat{\Sigma}_{sBL} - diag(\hat{\Sigma}_{sBL})$.

*Proof.* According to Equation 6 the covariance matrix reproduced from the SPFA factors is

$$\hat{\Sigma}_{SPFA} = \hat{\Lambda}_s (\mathbf{T}')^{-1} \hat{\Phi}_s (\mathbf{T})^{-1} \hat{\Lambda}'_s = \hat{\Lambda}_s \hat{\Lambda}_s = \hat{\Lambda}_{os} (\hat{\Lambda}'_{os} \mathbf{S}^{-1} \hat{\Lambda}_{os})^{-1} \hat{\Lambda}'_{os}. \tag{10}$$

The covariance matrix reproduced from factor score predictors can be computed from

$$\hat{\Sigma}_r = \mathbf{S B} (\mathbf{B'SB})^{-1} \mathbf{B'S}, \tag{11}$$

where **B** represents the weights for the factor score predictor (Beauducel & Hilger, 2015, 2019).

Entering $\mathbf{B}_{sBL} = \mathbf{S}^{-1} \hat{\Lambda}_{os} (\hat{\Lambda}'_{os} \mathbf{S}^{-1} \hat{\Lambda}_{os})^{-1/2} (\mathbf{T}^{-1})' \hat{\Phi}_s$ for **B** into Equation 11 and some transformation yields

$$\hat{\Sigma}_{sBL} = \hat{\Lambda}_{os} (\hat{\Lambda}'_{os} \mathbf{S}^{-1} \hat{\Lambda}_{os})^{-1/2} (\mathbf{T}^{-1})' \hat{\Phi}_s (\hat{\Phi}_s \mathbf{T}^{-1} (\mathbf{T}^{-1})' \hat{\Phi}_s)^{-1} \hat{\Phi}_s \mathbf{T}^{-1} (\hat{\Lambda}'_{os} \mathbf{S}^{-1} \hat{\Lambda}_{os})^{-1/2} \hat{\Lambda}'_{os}$$
$$= \hat{\Lambda}_{os} (\hat{\Lambda}'_{os} \mathbf{S}^{-1} \hat{\Lambda}_{os})^{-1} \hat{\Lambda}'_{os}. \tag{12}$$

This completes the Proof. □

It is notable that the identity of the non-diagonal elements of the covariance matrix reproduced from the SPFA factors and from the SPFA best linear factor score predictor does not depend on the amount of model error.

The equality of Anderson-Rubin's (1956) orthogonal factor score predictor with Takeuchi's factor score predictor $\hat{\mathbf{f}}_{Ta}$ has been shown in Beauducel (2015) so that it is sufficient to consider

$$\hat{\mathbf{f}}_{Ta} = (\mathbf{\Lambda}' \mathbf{\Sigma}^{-1} \mathbf{\Lambda})^{-1/2} \mathbf{\Lambda}' \mathbf{\Sigma}^{-1} \mathbf{x}. \tag{13}$$

As an orthogonal factor score predictor only makes sense for orthogonal factors, the orthogonal SPFA loadings $\hat{\Lambda}_{os} (\hat{\Lambda}'_{os} \mathbf{S}^{-1} \hat{\Lambda}_{os})^{-1/2}$ are inserted for $\mathbf{\Lambda}$ and **S** for $\mathbf{\Sigma}$ into equation 13. This yields

$$\hat{\mathbf{f}}_{sTa} = ((\hat{\Lambda}'_{os} \mathbf{S}^{-1} \hat{\Lambda}_{os})^{-1/2} \hat{\Lambda}'_{os} \mathbf{S}^{-1} \hat{\Lambda}_{os} (\hat{\Lambda}'_{os} \mathbf{S}^{-1} \hat{\Lambda}_{os})^{-1/2})^{-1/2} (\hat{\Lambda}'_{os} \mathbf{S}^{-1} \hat{\Lambda}_{os})^{-1/2} \hat{\Lambda}'_{os} \mathbf{S}^{-1} \mathbf{x}$$
$$= (\hat{\Lambda}'_{os} \mathbf{S}^{-1} \hat{\Lambda}_{os})^{-1/2} \hat{\Lambda}'_{os} \mathbf{S}^{-1} \mathbf{x}. \tag{14}$$

It follows from Equation 14 that $\hat{\mathbf{f}}_{sTa} = \hat{\mathbf{f}}_{sBL}$ for $\mathbf{T} = \mathbf{I}$ and $\hat{\Phi}_s = \mathbf{I}$ so that $Cov(\hat{\mathbf{f}}_{sTa}, \hat{\mathbf{f}}_{sTa}) = \mathbf{I}$, $Var(\hat{\mathbf{f}}_{sTa}) = \mathbf{I}$ and $Cor(\hat{\mathbf{f}}_{sTa}, \mathbf{f}_s) = \mathbf{I}$ for orthogonal SPFA factors and that Anderson-



Rubin's/Takeuchi's factor score predictor has perfect determinacy, is correlation preserving, conditionally unbiased, and has structural similarity, when the SPFA factors are orthogonal.

Krijnen, Wansbeek, and ten Berge (1996) proposed the best linear conditionally unbiased predictor, which can be written as

$$\hat{\mathbf{f}}_{Kr} = (\mathbf{\Lambda}'\mathbf{\Sigma}^{-1}\mathbf{\Lambda})^{-1}\mathbf{\Lambda}'\mathbf{\Sigma}^{-1}\mathbf{x}, \tag{15}$$

if $\mathbf{S}$ is invertible. Inserting $\hat{\mathbf{\Lambda}}_{os}(\hat{\mathbf{\Lambda}}_{os}'\mathbf{S}^{-1}\hat{\mathbf{\Lambda}}_{os})^{-1/2}(\mathbf{T}')^{-1}$ for $\mathbf{\Lambda}$ and $\mathbf{S}$ for $\mathbf{\Sigma}$ into equation 15 yields

$$\hat{\mathbf{f}}_{sKr} = (\mathbf{T}^{-1}(\hat{\mathbf{\Lambda}}_{os}'\mathbf{S}^{-1}\hat{\mathbf{\Lambda}}_{os})^{-1/2}\hat{\mathbf{\Lambda}}_{os}'\mathbf{S}^{-1}\hat{\mathbf{\Lambda}}_{os}(\hat{\mathbf{\Lambda}}_{os}'\mathbf{S}^{-1}\hat{\mathbf{\Lambda}}_{os})^{-1/2}(\mathbf{T}')^{-1})^{-1}\mathbf{T}^{-1}(\hat{\mathbf{\Lambda}}_{os}'\mathbf{S}^{-1}\hat{\mathbf{\Lambda}}_{os})^{-1/2}\hat{\mathbf{\Lambda}}_{os}'\mathbf{S}^{-1}\mathbf{x}$$
$$= \mathbf{T}'(\hat{\mathbf{\Lambda}}_{os}'\mathbf{S}^{-1}\hat{\mathbf{\Lambda}}_{os})^{-1/2}\hat{\mathbf{\Lambda}}_{os}'\mathbf{S}^{-1}\mathbf{x}. \tag{16}$$

It follows from Equations 7, 16, and from $\hat{\mathbf{\Phi}}_s\mathbf{T}^{-1} = \mathbf{T}'$ that $\hat{\mathbf{f}}_{sKr} = \hat{\mathbf{f}}_{sBL}$. which implies that Krijnen et al.'s conditionally unbiased predictor has perfect determinacy, is correlation preserving, conditionally unbiased, and has structural similarity.

Inserting $\hat{\mathbf{\Lambda}}_{os}(\hat{\mathbf{\Lambda}}_{os}'\mathbf{S}^{-1}\hat{\mathbf{\Lambda}}_{os})^{-1/2}(\mathbf{T}')^{-1}$ for $\mathbf{\Lambda}$ and $\hat{\mathbf{\Psi}}_{os}$ for $\mathbf{\Psi}$ into $\hat{\mathbf{f}}_{Ba} = (\mathbf{\Lambda}'\mathbf{\Psi}^{-2}\mathbf{\Lambda})^{-1}\mathbf{\Lambda}'\mathbf{\Psi}^{-2}\mathbf{x}$ for Bartlett's (1937) conditionally unbiased predictor and some transformation yields

$$\hat{\mathbf{f}}_{sBa} = \mathbf{T}'(\hat{\mathbf{\Lambda}}_{os}'\mathbf{S}^{-1}\hat{\mathbf{\Lambda}}_{os})^{1/2}(\hat{\mathbf{\Lambda}}_{os}'\hat{\mathbf{\Psi}}_{os}^{-2}\hat{\mathbf{\Lambda}}_{os})^{-1}\hat{\mathbf{\Lambda}}_{os}'\hat{\mathbf{\Psi}}_{os}^{-2}\mathbf{x} \tag{17}$$

so that

$$Cov(\hat{\mathbf{f}}_{sBa}, \hat{\mathbf{f}}_{sBa}) = \mathbf{T}'(\hat{\mathbf{\Lambda}}_{os}'\mathbf{S}^{-1}\hat{\mathbf{\Lambda}}_{os})^{1/2}(\hat{\mathbf{\Lambda}}_{os}'\hat{\mathbf{\Psi}}_{os}^{-2}\hat{\mathbf{\Lambda}}_{os})^{-1}\hat{\mathbf{\Lambda}}_{os}'\hat{\mathbf{\Psi}}_{os}^{-2}\mathbf{S}\hat{\mathbf{\Psi}}_{os}^{-2}\hat{\mathbf{\Lambda}}_{os}(\hat{\mathbf{\Lambda}}_{os}'\hat{\mathbf{\Psi}}_{os}^{-2}\hat{\mathbf{\Lambda}}_{os})^{-1}(\hat{\mathbf{\Lambda}}_{os}'\mathbf{S}^{-1}\hat{\mathbf{\Lambda}}_{os})^{1/2}\mathbf{T}. \tag{18}$$

Inserting $\hat{\mathbf{\Psi}}_{os}^{-2}\hat{\mathbf{\Lambda}}_{os} = \mathbf{S}^{-1}\hat{\mathbf{\Lambda}}_{os}(\mathbf{I} + \mathbf{\Phi}_s\hat{\mathbf{\Lambda}}_{os}'\hat{\mathbf{\Psi}}_{os}^{-2}\hat{\mathbf{\Lambda}}_{os})$ and $\hat{\mathbf{\Psi}}_{os}^{-2}\hat{\mathbf{\Lambda}}_{os}(\mathbf{I} + \mathbf{\Phi}_s\hat{\mathbf{\Lambda}}_{os}'\hat{\mathbf{\Psi}}_{os}^{-2}\hat{\mathbf{\Lambda}}_{os})^{-1} = \mathbf{S}^{-1}\hat{\mathbf{\Lambda}}_{os}$ according to Jöreskog (1969, Eq. 10) and some transformation yields

$$Cov(\hat{\mathbf{f}}_{sBa}, \hat{\mathbf{f}}_{sBa}) = \mathbf{T}'((\hat{\mathbf{\Lambda}}_{os}'\hat{\mathbf{\Psi}}_{os}^{-2}\hat{\mathbf{\Lambda}}_{os})^{-1} + \mathbf{\Phi}_s)((\hat{\mathbf{\Lambda}}_{os}'\hat{\mathbf{\Psi}}_{os}^{-2}\hat{\mathbf{\Lambda}}_{os})^{-1} + \mathbf{\Phi}_s)((\hat{\mathbf{\Lambda}}_{os}'\hat{\mathbf{\Psi}}_{os}^{-2}\hat{\mathbf{\Lambda}}_{os})^{-1} + \mathbf{\Phi}_s)^{-1/2}\mathbf{T} = \mathbf{\Phi}_s, \tag{19}$$

so that $Var(\hat{\mathbf{f}}_{sBa}) = \mathbf{I}$. It follows from $Var(\hat{\mathbf{f}}_{sBL}) = Var(\hat{\mathbf{f}}_{sKr}) = \mathbf{I}$ that

$$Cor(\hat{\mathbf{f}}_{sBa}, \hat{\mathbf{f}}_{sKr}) = \mathbf{T}'(\hat{\mathbf{\Lambda}}_{os}'\mathbf{S}^{-1}\hat{\mathbf{\Lambda}}_{os})^{1/2}(\hat{\mathbf{\Lambda}}_{os}'\hat{\mathbf{\Psi}}_{os}^{-2}\hat{\mathbf{\Lambda}}_{os})^{-1}\hat{\mathbf{\Lambda}}_{os}'\hat{\mathbf{\Psi}}_{os}^{-2}\mathbf{S}\mathbf{S}^{-1}\hat{\mathbf{\Lambda}}_{os}(\hat{\mathbf{\Lambda}}_{os}'\mathbf{S}^{-1}\hat{\mathbf{\Lambda}}_{os})^{-1/2}\mathbf{T} = \hat{\mathbf{\Phi}}_s, \tag{20}$$

that $Cor(\hat{\mathbf{f}}_{sBa}, \mathbf{f}_s) = \hat{\mathbf{\Phi}}_s$ and that $\hat{\mathbf{\Sigma}}_{sBa} = \hat{\mathbf{\Sigma}}_{sKr} = \hat{\mathbf{\Sigma}}_{SPFA}$.

Inserting $\hat{\mathbf{\Lambda}}_s(\mathbf{T}')^{-1}$ for $\mathbf{\Lambda}$ into Harman's (1976) ideal variable score predictor given by $\mathbf{f}_{Ha} = (\mathbf{\Lambda}'\mathbf{\Lambda})^{-1}\mathbf{\Lambda}'\mathbf{x}$ yields

$$\hat{\mathbf{f}}_{sHa} = (\mathbf{T}^{-1}\hat{\mathbf{\Lambda}}_s'\hat{\mathbf{\Lambda}}_s(\mathbf{T}')^{-1})^{-1}\mathbf{T}^{-1}\hat{\mathbf{\Lambda}}_s'\mathbf{x} = \mathbf{T}'(\hat{\mathbf{\Lambda}}_s'\hat{\mathbf{\Lambda}}_s)^{-1}\hat{\mathbf{\Lambda}}_s'\mathbf{x}, \tag{21}$$

so that

$$Cov(\hat{\mathbf{f}}_{sHa}, \hat{\mathbf{f}}_{sHa}) = \mathbf{T}'(\hat{\mathbf{\Lambda}}_s'\hat{\mathbf{\Lambda}}_s)^{-1}\hat{\mathbf{\Lambda}}_s'\mathbf{S}\hat{\mathbf{\Lambda}}_s(\hat{\mathbf{\Lambda}}_s'\hat{\mathbf{\Lambda}}_s)^{-1}\mathbf{T}. \tag{22}$$

It follows from Equation 6 that



$$Var(\hat{\mathbf{f}}_{sHa}) = diag((\mathbf{T}^{-1}\hat{\mathbf{\Lambda}}_s'\hat{\mathbf{\Lambda}}_s(\mathbf{T}')^{-1})^{-1}\mathbf{T}^{-1}\hat{\mathbf{\Lambda}}_s'\mathbf{S}\hat{\mathbf{\Lambda}}_s(\mathbf{T}')^{-1}(\mathbf{T}^{-1}\hat{\mathbf{\Lambda}}_s'\hat{\mathbf{\Lambda}}_s(\mathbf{T}')^{-1})^{-1})$$
$$= diag(\hat{\mathbf{\Phi}}_s + \mathbf{T}'(\hat{\mathbf{\Lambda}}_s'\hat{\mathbf{\Lambda}}_s)^{-1}\hat{\mathbf{\Lambda}}_s'(\mathbf{\Psi}_s^2 + \mathbf{\Omega}_s)\hat{\mathbf{\Lambda}}_s(\hat{\mathbf{\Lambda}}_s'\hat{\mathbf{\Lambda}}_s)^{-1}\mathbf{T}).$$

(23)

From $\hat{\mathbf{\Lambda}}_s = \mathbf{v}\mathbf{e}^{1/2}$, with eigenvector $\mathbf{v}$, $\mathbf{v}'\mathbf{v} = \mathbf{I}$, and the diagonal matrix of eigenvalues $\mathbf{e} > \mathbf{0}$,

follows $Var(\hat{\mathbf{f}}_{sHa}) = diag(\hat{\mathbf{\Phi}}_s + \mathbf{T}'\mathbf{e}^{-1/2}\mathbf{v}_s'(\mathbf{\Psi}_s^2 + \mathbf{\Omega}_s)\mathbf{v}_s\mathbf{e}^{-1/2}\mathbf{T})$. If not all the variance in $\mathbf{S}$ is

explained by the SPFA factors, which is rather likely, there will be positive eigenvalues of

$\mathbf{\Psi}_s^2 + \mathbf{\Omega}_s$ so that $Var(\hat{\mathbf{f}}_{sHa}) \geq \mathbf{I}$. Since Equations 5 and 21 imply

$$Cov(\hat{\mathbf{f}}_{sHa}, \mathbf{f}_s) = \mathbf{T}'(\hat{\mathbf{\Lambda}}_s'\hat{\mathbf{\Lambda}}_s)^{-1}\hat{\mathbf{\Lambda}}_s'\hat{\mathbf{\Lambda}}_s(\mathbf{T}')^{-1}\hat{\mathbf{\Phi}}_s = \hat{\mathbf{\Phi}}_s,$$

(24)

it follows from $Var(\hat{\mathbf{f}}_{sHa}) \geq \mathbf{I}$ that $diag(Cor(\hat{\mathbf{f}}_{sHa}, \mathbf{f}_s)) \leq \mathbf{I}$. Thus, the determinacy of Harman's

score predictor is not perfect. It also follows from $diag(\hat{\mathbf{\Phi}}_s + \mathbf{T}'\mathbf{e}^{-1/2}\mathbf{v}_s'(\mathbf{\Psi}_s^2 + \mathbf{\Omega}_s)\mathbf{v}_s\mathbf{e}^{-1/2}\mathbf{T}) \geq \mathbf{I}$

that the absolute size of the non-diagonal elements of

$diag(\hat{\mathbf{\Phi}}_s + \mathbf{T}'\mathbf{e}^{-1/2}\mathbf{v}_s'(\mathbf{\Psi}_s^2 + \mathbf{\Omega}_s)\mathbf{v}_s\mathbf{e}^{-1/2}\mathbf{T})^{-1/2}\hat{\mathbf{\Phi}}_s$ is smaller than the absolute size of the non-

diagonal elements of $\hat{\mathbf{\Phi}}_s$. Thus, Harman's score predictor is not correlation-preserving. With

respect to conditional unbiasedness it should be noted that Harman's factor score predictor

correlate to a lower degree with the other factors than the factors themselves do. Entering the

weights of SPFA Harman's factor score predictor $\mathbf{B}_{sHa} = \hat{\mathbf{\Lambda}}_s(\hat{\mathbf{\Lambda}}_s'\hat{\mathbf{\Lambda}}_s)^{-1}\mathbf{T}$ into Equation 11 and

some transformation yields $\hat{\mathbf{\Sigma}}_{sHa} = \mathbf{S}\hat{\mathbf{\Lambda}}_s(\hat{\mathbf{\Lambda}}_s'\mathbf{S}\hat{\mathbf{\Lambda}}_s)^{-1}\hat{\mathbf{\Lambda}}_s'\mathbf{S} = \mathbf{S}\hat{\mathbf{\Lambda}}_{os}(\hat{\mathbf{\Lambda}}_{os}'\mathbf{S}\hat{\mathbf{\Lambda}}_s)^{-1}\hat{\mathbf{\Lambda}}_{os}'\mathbf{S}$, indicating that

the SPFA Harman factor score predictor has no structural similarity with the SPFA factors.

## Models with a single large loading per factor

The performance of SPFA and CFM to identify factors with a single large loading was

compared by means of a simulation study. The conditions of the simulation study were salient

loading size ($sl$ = .50, .60, .70, .80), sample size ($n$ = 200, 400, 1,000), and number of

orthogonal factors ($q$ = 2, 5, 8). Each factor was defined by one salient population loading and

by two small population loadings of .30. In order to investigate single-item identification in a

context of a large number of irrelevant variables, seven population loadings per factor were

zero. For $q$ = 2 this results in 14 variables with zero loadings and 6 variables with non-zero

loadings (see Table 1).



Table 1. Population loadings for $q = 2$ and $sl = .70$

| Variable | F1 | F2 |
|---|---|---|
| X1 | **.70** | .00 |
| X2 | **.30** | .00 |
| X3 | **.30** | .00 |
| X4 | .00 | .00 |
| X5 | .00 | .00 |
| X6 | .00 | .00 |
| X7 | .00 | .00 |
| X8 | .00 | .00 |
| X9 | .00 | .00 |
| X10 | .00 | .00 |
| X11 | .00 | **.70** |
| X12 | .00 | **.30** |
| X13 | .00 | **.30** |
| X14 | .00 | .00 |
| X15 | .00 | .00 |
| X16 | .00 | .00 |
| X17 | .00 | .00 |
| X18 | .00 | .00 |
| X19 | .00 | .00 |
| X20 | .00 | .00 |

*Note.* Loadings ≥ .30 are given in bold face.

Three rotation methods were investigated (Varimax, Parsimax, Infomax). Varimax-rotation (Kaiser, 1958) was investigated because it is still one of the most popular rotation methods (Weide & Beauducel, 2019), Parsimax-rotation (Crawford & Ferguson, 1970) was investigated because it can be regarded as a method where the simple structure across rows and columns of a loading matrix is balanced (Browne, 2001). Infomax-rotation (McKeon, 1968) was investigated in order to enhance the heterogeneity of rotation methods. Overall, the design of the simulation study comprises 4 ($sl$) x 3 ($q$) x 3 ($n$) = 36 conditions. For each condition 1,000 samples were generated. Data generation and analysis was based on IBM SPSS Version 26. Common factor scores **f** and unique factor scores **u** were generated from normal distributions with $\mu = 0$ and $\sigma = 1$ by the method of Box and Muller (1958) from uniformly distributed numbers generated by the Mersenne twister integrated in SPSS. Observed variable scores **x** were computed from the population loadings, the corresponding unique loadings, and the common and unique factor scores according to Equation 1 and submitted to CFM and SPFA. The factor rotation was performed according to Bernaards and Jennrich (2005). The 36,000 CFM loading patterns and the 36,000 SPFA loading patterns were submitted to Varimax, Parsimax, Infomax and Target rotation. One dependent variable



is the mean congruence coefficient (Tucker, 1951) describing the mean similarity of the sample factors with the population factors for each condition, extraction, and rotation method. As a second dependent variable, the percentage of selection of the variable with the largest population loading from the sample loading matrix will also be investigated. The absolute sample loading of the variable that might be selected as a single-item indicator should be substantially larger than the sample loading of the other variables on the respective factor. Moreover, the sample loading of the variable should also be substantially larger than the absolute sample loading of the respective variable on the remaining factors. Therefore, the percentage of absolute sample loadings that are at least .05 (and in a second condition at least .10) greater than the remaining absolute sample loadings (in rows and columns of the loading pattern) for variables with the salient population loading on a factor ($sl$), is investigated. Since the population salient loading per factor is at least .20 greater than the small population loading (for $sl = .50$), this percentage is termed the percentage of correctly identified single-item indicators. As a benchmark, the mean congruence for Target-rotation towards the population loading matrix was also computed.

The results are given in Figure 1 and can be summarized as follows: For $n \geq 400$ and $q = 8$, the mean congruence with the population loadings was greater for SPFA loadings than for CFM loadings for all salient loading sizes. For $sl \geq .60$ and $q \geq 5$ the mean congruence with the population loadings was greater for SPFA loadings than for CFM loadings for all sample sizes. For $q = 2$ and $sl \leq .60$ the mean congruence with the population loadings was greater for CFM loadings than for SPFA loadings. For $sl = .80$ the mean congruence with the population loadings was greater for SPFA loadings than for CFM loadings for all numbers of factors and all sample sizes. The effect of the factor rotation method was negligible. To sum up, when the number of factors was small and the salient loading size was moderate to small, CFM had larger mean congruence than SPFA. With five or more factors and moderate to large loadings SPFA had larger mean congruence than CFM. For large loadings SPFA had larger mean congruence than CFM for all numbers of factors and sample sizes that were investigated.



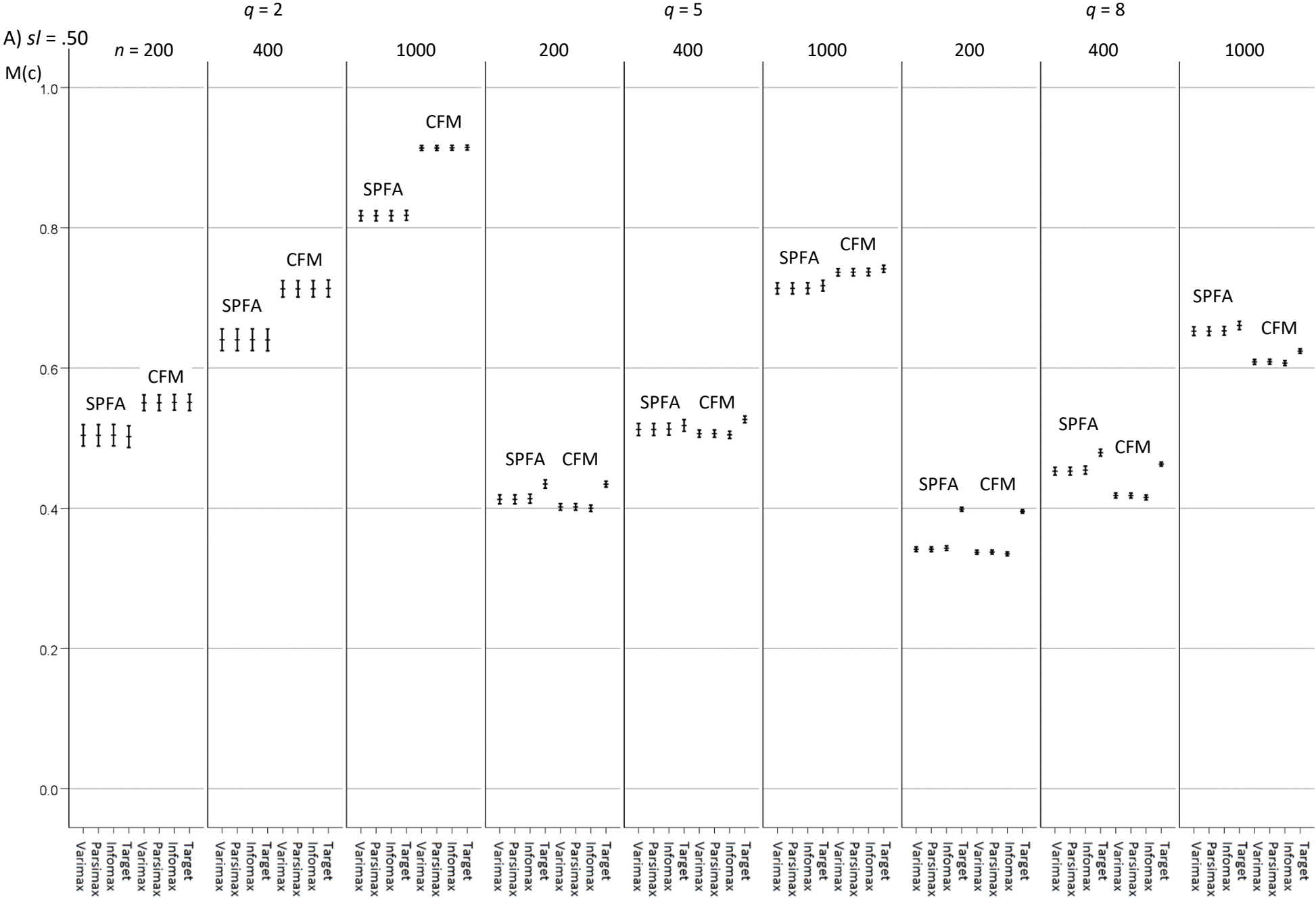





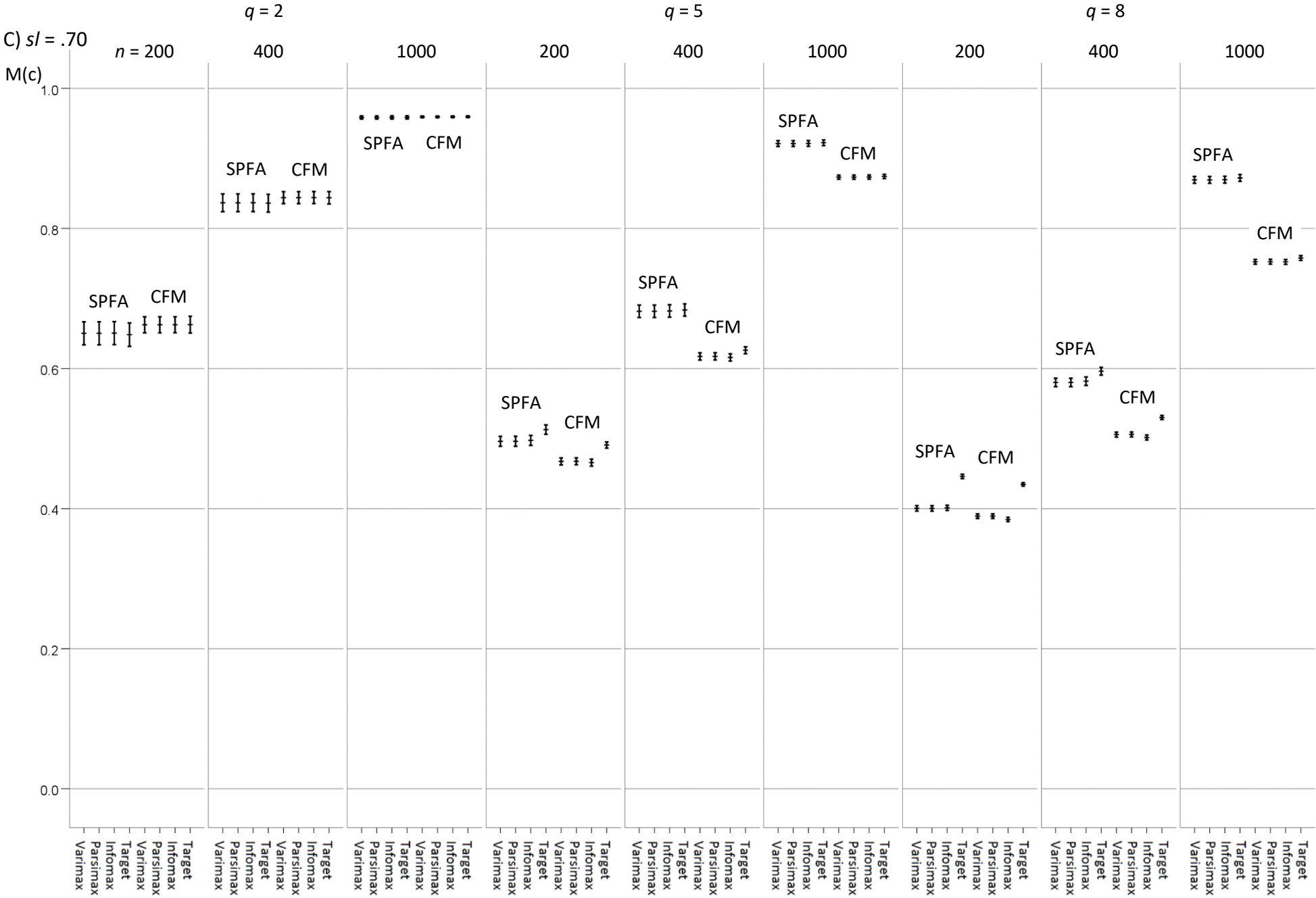



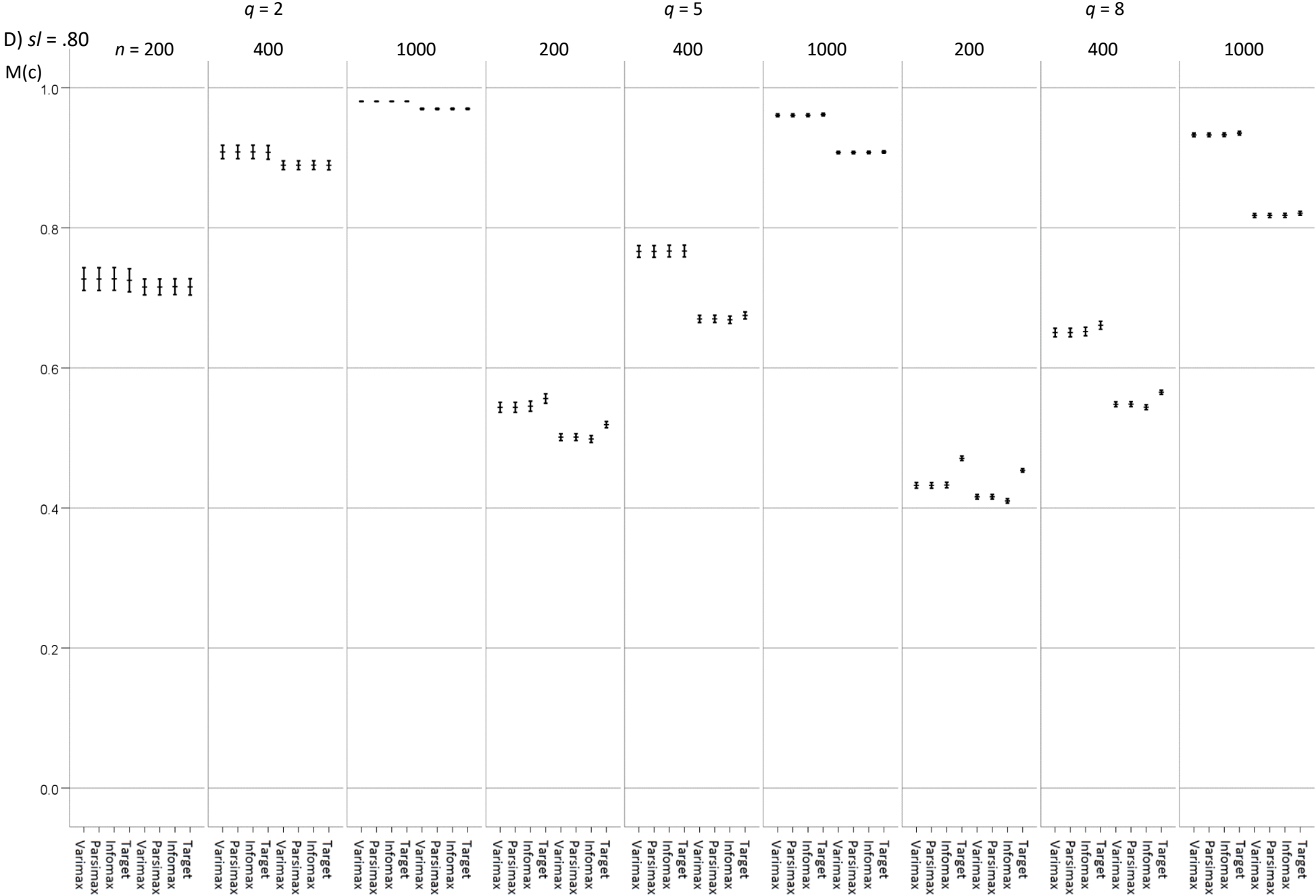

Figure 1. Mean congruences M(c) for *l* = .50 (A), *l* = .60 (B), *l* = .70 (C), *l* = .80 (D).



Since the results for the mean congruence coefficients were nearly identical for the different methods of factor rotation, the percentage of correctly identified single-item indicators was only investigated for the Varimax rotation. Overall, the percentage of correctly identified single-item indicators was larger for SPFA than for CFM (see Table 2). The difference between SPFA and CFM was larger for larger sample sizes. Even for $q = 2$ and $sl \leq$ .60, when the mean congruence with the population loadings was greater for CFM loadings than for SPFA loadings, the percentage of correctly identified single-item indicators was greater for SPFA than for CFM.

Table 2. Percent of largest population loadings detected in Varimax rotation sample patterns (separate for *n*, across all *q* and *sl*)

| *sl* | *q* | *n* | CFM .05 | SPFA .05 | CFM .10 | SPFA .10 |
|------|-----|------|---------|----------|---------|----------|
| .50  | 2   | 200  | 22.80   | 28.25    | 17.65   | 28.25    |
| .50  | 2   | 400  | 38.15   | 45.50    | 29.80   | 45.50    |
| .50  | 2   | 1000 | 66.15   | 76.20    | 53.10   | 76.20    |
| .50  | 5   | 200  | 13.26   | 18.78    | 9.44    | 18.10    |
| .50  | 5   | 400  | 22.48   | 31.86    | 15.94   | 31.82    |
| .50  | 5   | 1000 | 44.10   | 59.84    | 30.38   | 59.84    |
| .50  | 8   | 200  | 12.04   | 14.20    | 8.34    | 12.50    |
| .50  | 8   | 400  | 18.13   | 23.99    | 12.43   | 23.31    |
| .50  | 8   | 1000 | 34.38   | 52.34    | 20.61   | 52.34    |
| .60  | 2   | 200  | 28.85   | 36.45    | 23.60   | 36.45    |
| .60  | 2   | 400  | 52.80   | 61.75    | 44.15   | 61.75    |
| .60  | 2   | 1000 | 88.05   | 93.50    | 77.95   | 93.50    |
| .60  | 5   | 200  | 17.96   | 24.08    | 12.60   | 23.28    |
| .60  | 5   | 400  | 31.02   | 43.20    | 22.64   | 43.16    |
| .60  | 5   | 1000 | 67.02   | 79.28    | 52.02   | 79.28    |
| .60  | 8   | 200  | 15.35   | 18.30    | 10.95   | 16.31    |
| .60  | 8   | 400  | 24.25   | 32.71    | 16.40   | 31.96    |
| .60  | 8   | 1000 | 51.39   | 69.86    | 34.43   | 69.86    |



Table 2. (continued)

| sl | q | n | CFM .05 | SPFA .05 | CFM .10 | SPFA .10 |
|----|---|---|---------|----------|---------|----------|
| .70 | 2 | 200 | 38.50 | 46.85 | 31.95 | 46.85 |
| .70 | 2 | 400 | 69.00 | 75.15 | 60.55 | 75.15 |
| .70 | 2 | 1000 | 97.40 | 98.80 | 93.75 | 98.80 |
| .70 | 5 | 200 | 22.76 | 30.32 | 16.68 | 29.44 |
| .70 | 5 | 400 | 42.60 | 55.28 | 31.62 | 55.26 |
| .70 | 5 | 1000 | 87.62 | 93.32 | 76.28 | 93.32 |
| .70 | 8 | 200 | 19.96 | 23.10 | 14.36 | 20.51 |
| .70 | 8 | 400 | 32.56 | 43.60 | 21.81 | 42.70 |
| .70 | 8 | 1000 | 71.81 | 86.31 | 55.03 | 86.31 |
| .80 | 2 | 200 | 50.95 | 59.00 | 42.55 | 59.00 |
| .80 | 2 | 400 | 82.70 | 87.20 | 76.50 | 87.20 |
| .80 | 2 | 1000 | 99.85 | 100.00 | 98.70 | 100.00 |
| .80 | 5 | 200 | 30.14 | 38.78 | 22.04 | 37.90 |
| .80 | 5 | 400 | 55.86 | 69.26 | 43.64 | 69.26 |
| .80 | 5 | 1000 | 96.88 | 98.68 | 92.10 | 98.68 |
| .80 | 8 | 200 | 24.96 | 29.11 | 17.93 | 25.61 |
| .80 | 8 | 400 | 42.88 | 55.33 | 29.59 | 54.75 |
| .80 | 8 | 1000 | 88.85 | 96.14 | 77.34 | 96.14 |

*Note.* CFM .05/SPFA .05: Percentage of absolute sample loadings for variables with the largest population loading that are greater by .05 or more than the second largest loading in the column and row of the loading matrix. CFM .10/SPFA .10: Percentage of absolute sample loadings for variables with the largest population loading that are greater by .10 or more than the second largest loading in the column and row of the loading matrix.

## Discussion

The present study evaluates SPFA, which has initially been developed in order to provide optimal factor score predictors in the context of factor analysis (Beauducel & Hilger, 2019), as a method in its own rights. Therefore, the quality of SPFA factor score predictors as measures for the SPFA factors was evaluated algebraically. It turns out that the SPFA best linear predictor has a perfect correlation with the corresponding SPFA factor (determinacy),



that it is correlation preserving, and has structural similarity. Thus, the SPFA factor score predictor can be used as an optimal measure for the SPFA factor. It was also found that in the context of SPFA, Krijnen et al.'s (1996) and Bartlett's (1937) conditionally unbiased predictors are identical with the SPFA best linear predictor. Takeuchi's orthogonal factor score predictor is identical with the SPFA best linear predictor when the SPFA factors are orthogonal. In contrast, Harman's (1976) ideal variable factor score predictor does not correlate perfectly with the SPFA factors and has no structural similarity.

The perfect determinacy of the SPFA best linear predictor and its perfect correlation with the conditional unbiased predictors provided by Krijnen et al. (1996) and Bartlett (1937) underlines that SPFA considered in its own right is a regression component model in the sense of Schönemann and Steiger (1976). However, it is also shown that this does not imply that any factor score predictor known from the context of factor analysis has a perfect determinacy. Since Harman's ideal variable factor score predictor has no perfect determinacy it is recommended to calculate the best linear predictor in the context of SPFA when individual scores of the participants are needed.

The utility of SPFA and CFM for identification of factors based on very few variables was compared by means of a simulation study. Although quite different methods of factor rotation were investigated the effect of Varimax-, Parsimax- or Infomax-rotation on mean congruence with the population loadings was negligible. This indicates that the choice of rotation method is not an issue when factors with very few salient loadings are to be identified. Probably, the issue of factor rotation is more important when the pattern of population loadings is more complex. Moreover, the results based on coefficients of congruence reveal that CFM can be used for the identification of factors with very few salient loadings or for a variable that can be used as a single-item indicator when only two factors and moderate loadings are considered. With five or more factors and with moderate to large salient loadings, SPFA should be preferred as a method for the identification of factors based on very few salient loadings. However, when relative loading thresholds were used for the identification of single-item indicators, it was found that SPFA correctly identified a substantially larger percentage of single-item indicators than CFM in all conditions, even in



conditions with a small number of factors and moderate loadings. To sum up, the results of the simulation study reveal that SPFA can be recommended as a method for the identification of single-item indicators. An R-code for SPFA is given in the supplement of Beauducel and Hilger (2019).



References


Anderson, T.W., & Rubin, H. (1956). Statistical inference in factor analysis. *Proceedings of the Third Berkeley Symposium of Mathematical Statistics and Probability, 5*, 111–150.

Bartlett, M. S. (1937). The statistical conception of mental factors. *British Journal of Psychology, 28*, 97–104. doi:10.1111/j.2044-8295.1937.tb00863.x

Beauducel, A. (2005). How to describe the difference between factors and corresponding factor score estimates. *Methodology, 1*, 143-158. doi: 10.1027/1614-2241.1.4.143

Beauducel, A. (2007). In spite of indeterminacy many common factor score estimates yield an identical reproduced covariance matrix. *Psychometrika, 72*, 437–441. doi:10.1007/s11336-005-1467-5

Beauducel, A. & Hilger, N. (2015). Extending the debate between Spearman and Wilson 1929: When do single variables optimally reproduce the common part of the observed covariances? *Multivariate Behavioral Research, 50*, 555-567. doi: 10.1080/00273171.2015.1059311.

Beauducel, A. & Hilger, N. (2019). Score predictor factor analysis: Reproducing observed covariances by means of factor score predictors. *Frontiers in Psychology. Quantitative Psychology and Measurement,* 10: 1895. doi : 10.3389/fpsyg.2019.01895

Bergkvist, L. & Rossiter, J.R. (2007). The predictive validity of multiple-item versus single-item measures of the same constructs. *Journal of Marketing Research, 44*, 175–184. doi: 10.1509/jmkr.44.2.175

Bernaards, C. A., and Jennrich, R. I. (2005). Gradient projection algorithms and software for arbitrary rotation criteria in factor analysis. *Educational and Psycholological Measurement, 65*, 676–696. doi: 10.1177/0013164404272507

Box, G. E. P. & Muller, M. E. (1958). A note on the generation of random normal deviates. *Annals of Mathematical Statistics, 29*, 610–611. doi: 10.1214/aoms/1177706645

Browne, M. W. (2001). An overview of analytic rotation in exploratory factor analysis. *Multivariate Behavioral Research, 36*, 111–150. doi: 10.1207/S15327906MBR3601_05

Crawford, C. B. & Ferguson, G. A. (1970). A general rotation criterion and its use in orthogonal rotation. *Psychometrika, 35*, 321-332. doi: 10.1007/BF02310792

Gorsuch, R.L. (1983). *Factor Analysis*. Hillsdale, NJ: Lawrence Erlbaum.

Grice, J.W. (2001). Computing and evaluating factor scores. *Psychological Methods, 6*, 430–450. doi:10.1037/1082-989X.6.4.430

Guttman, L. (1955). The determinacy of factor score matrices with applications for five other problems of common factor theory. *British Journal of Statistical Psychology, 8*, 65–82. doi:10.1111/j.2044-8317.1955.tb00321.x

Harman, H. H. (1976). *Modern factor analysis* (3rd ed.). Chicago, IL: The University of Chicago Press.

Harman, H. H., & Jones, W. H. (1966). Factor analysis by minimizing residuals (minres). *Psychometrika, 31*, 351–368. doi:10.1007/BF02289468

Jöreskog, K.G. (1969). A general approach to confirmatory maximum likelihood factor analysis. *Psychometrika, 34*, 183–202. doi: 10.1007/BF02289343

Kaiser, H. F. (1958). The varimax criterion for analytic rotation in factor analysis. *Psychometrika, 23*, 187–200. doi: 10.1007/BF02289233

MacCallum, R. C., Tucker, L. R. (1991). Representing sources of error in the common-factor model: Implications for theory and practice. *Psychological Bulletin, 109*, 502–511. doi: 10.1037/0033-2909.109.3.502

McDonald, R. P. (1981). Constrained least squares estimators of oblique common factors. *Psychometrika, 46*, 337–341. doi:10.1007/BF02293740





McKeon, J. J. (1968). *Rotation for maximum association between factors and tests.* Unpublished manuscript, Biometric Laboratory, George Washington University.

Nicewander, W. A. (2019). A perspective on the mathematical and psychometric aspects of factor indeterminacy. *Multivariate Behavioral Research.* doi:10.1080/00273171.2019.1684872

Rammstedt, B. & Beierlein, C. (2014). Can't we make it any shorter? The limits of personality assessment and ways to overcome them. *Journal of Individual Differences, 35*, 212-220. doi:10.1027/1614-0001/a000141

Schönemann, P. H., & Steiger, J. H. (1976). Regression component analysis. *British Journal of Mathemathical and Statistical Psychology, 29*, 175–189. doi: 10.1111/j.2044-8317.1976.tb00713.x

Thurstone, L. L. (1935). *The Vectors of mind.* Chicago, IL: University of Chicago Press.

Tucker, L. R. (1951). *A method for synthesis of factor analysis studies. Personnel Research Section Report No. 984.* Washington, D.C.: Department of the Army.

Weide, A. C. & Beauducel, A. (2019). Varimax rotation based on gradient projection is a feasible alternative to SPSS. *Frontiers in Psychology. Quantitative Psychology and Measurement,* 10: 645. doi: 10.3389/fpsyg.2019.00645

Ziegler, M., Kemper, Ch. J. & Kruyen, P. (2014). Short scales – five misunderstandings and ways to overcome them. *Journal of Individual Differences, 35*, 212-220. Doi:10.1027/1614-0001/a000148